\documentclass[oneside]{article}
\usepackage[T1]{fontenc}
\usepackage{authblk}
\usepackage{float}
\usepackage{amsmath}
\usepackage{amssymb}
\usepackage{graphicx}
\usepackage{xfrac}
\usepackage[english]{babel}
\usepackage{lmodern}
\usepackage[T1]{fontenc}
\usepackage[latin9]{inputenc}
\usepackage{mathtools}
\usepackage{graphicx}
\usepackage{esint}
\usepackage[unicode=true,
 bookmarks=false,
breaklinks=false,pdfborder={0 0 1},backref=false,colorlinks=false]
 {hyperref}
\usepackage[a4paper, total={210mm,297mm},margin=2.0cm]{geometry}

\usepackage{graphicx}
\usepackage{dcolumn}
\usepackage{bm}

\usepackage{mathptmx}
\usepackage{etoolbox}

\title{\textbf{Lightweight Lattice Boltzmann}}

\author[1]{A. Tiribocchi}

\author[2]{A. Montessori}

\author[3]{G. Amati}

\author[1]{M. Bernaschi}

\author[4]{F. Bonaccorso}

\author[3]{S. Orlandini}

\author[5,6,1]{S. Succi}

\author[1]{M. Lauricella \thanks{Electronic address: \texttt{marco.lauricella@cnr.it}; Corresponding author}}

\affil[1]{Istituto per le Applicazioni del Calcolo CNR, via dei Taurini 19, 00185 Rome, Italy}
\affil[2]{Department of Engineering, Roma Tre University, Via Vito Volterra 62, 00146 Rome, Italy}
\affil[3]{SCAI, SuperComputing Applications and Innovation Department, CINECA, Via dei Tizii, 6, Rome 00185, Italy}
\affil[4]{Department of Physics and INFN, University of Rome Tor Vergata, Via della Ricerca Scientifica 1, 00133 Rome, Italy}
\affil[5]{Center for Life Nano Science@La Sapienza, Istituto Italiano di Tecnologia, 00161 Roma, Italy}
\affil[6]{Department of Physics, Harvard University, Cambridge, MA, 02138, USA}

\date{\today}

\begin{document}

\maketitle

\begin{abstract}
A GPU-accelerated version of the lattice Boltzmann method for efficient simulation of soft materials is introduced. Unlike standard approaches, this method reconstructs the distribution functions from available hydrodynamic variables (density, momentum, and pressure tensor) without storing the full set of discrete populations. This scheme shows satisfactory numerical stability, significantly lower memory requirements,  and data access cost. A series of benchmark tests of relevance to soft matter, such as collisions of fluid droplets, is discussed to validate the method. The results can be of particular interest for high-performance simulations of soft matter systems on future exascale computers.
\end{abstract}

\maketitle

\section{\label{sec1}Introduction}

Minimizing the cost of accessing data is one of the most compelling 
challenges of present-day high-performance computing \cite{jamshed2015using}.
This is particularly true for memory-bound computational schemes used in computational fluid dynamics. Among them,
the lattice Boltzmann (LB) method \cite{succi2018} is characterized by a comparatively
low arithmetic intensity spanning in a range between 1 and 5 Flops/Byte ratio depending on the implementation, hence below the roofline threshold beyond which computing resources become the limiting factor \cite{exasc1,exasc2}.

Given the widespread use of LB across a broad spectrum of scales and
regimes, it is of prime interest to develop techniques to mitigate the
impact of data access on future (exascale) LB simulations. Several such techniques have been developed in the past, based on hierarchical memory access \cite{lamura,mattila}, data organization in terms of
array of structures and structure of arrays \cite{tripiccione}, as well as special
crystallographic lattices which double the resolution at a given
level of memory occupation \cite{namburi}. Further, several approaches have been proposed to reduce memory occupancy that can be summarized in two main strategies: the first focuses on algorithmic implementation, while the second exploits different numerical representations. As a remarkable example of the algorithmic approach, we recall the so-called ``in-place'' streaming, like esoteric pull \cite{lehmann2022esoteric,geier2017esoteric}, while in the latest approach low precision and shifting equilibrium distribution are used with single precision instead double or with a mixed single precision/half-precision representation, with no significant loss in the quality of the simulation but reducing by a factor two total memory occupancy \cite{lehmann2022accuracy,gray2016enhancing}.

In this work, we present a new strategy to implement the LB approach based on the idea of refraining from storing the populations and reconstructing them on the fly, based on the available hydrodynamic information. This approach grants a significant gain in memory usage since it avoids population storage (in the following, simply called ``without populations''). Further, this is particularly useful for  multi-component and
multiple species applications in which
each species is carried by a common flow field, thereby requiring
a single hydrodynamic field (density) as opposed to the full kinetic
representation with $O(30)$ populations. 
A similar argument applies to flows far from equilibrium, or relativistic hydrodynamics \cite{rezzolla,rezzolla2}, which demand higher order lattices sometimes entailing
hundreds of discrete populations per species.   

To validate the accuracy of the method, the present approach is tested on a selection of physically relevant situations in soft matter. More specifically, we simulate a 3D off-axis and a 3D head-on collision between two 
equally sized fluid droplets in which the merging is suppressed plus an off-axis collision in which coalescence occurs. These simulations are compared with those performed using an LB version with populations, and the accuracy of the results is quantitatively assessed by computing the time evolution of the droplet densities and the relative error. Even though the reconstruction procedure 
suffers from an inherent approximation with respect to the standard implementation of the LB method (see Section II), our results show that the error remains negligible, the numerical accuracy is generally preserved and the physics is correctly reproduced. Compared to previous models \cite{latt2006,chen2017}, the distinctive feature of the reconstruction procedure discussed here is that, unlike a standard LB approach, it considerably enhances the computational performance of GPU-based lattice Boltzmann codes, entailing about 40 percent of memory savings on large scale simulations. This represents a significant step  towards the study of the multiscale physics of soft materials on modern  GPU-accelerated machines, where the minimization of the burden imposed by memory requirements and data access is often mandatory.

The paper is structured as follows. In section II we illustrate the idea behind the LB without populations, while in section III we shortly describe the implementation and performance data. In section IV we present some test cases which validate the method and in the last section we discuss possible perspectives.

\section{\label{sec12}Lattice Boltzmann without populations: the basic idea}

The LB scheme is based on a set of discrete populations 
$f_i(\vec{x},t) = f(\vec{x},\vec{\upsilon },t) \delta(\vec{v}-\vec{c}_i)$ describing the probability 
of finding a representative fluid particle at position $\vec{x}$
and time $t$ with a discrete molecular velocity $\vec{\upsilon }=\vec{c}_i$.
The discrete velocities are chosen according to suitable crystallographic
symmetries ensuring compliance with the basic conservation laws of 
macroscopic fluids.
This implies a set of order $b \sim 20-30$ discrete 
populations $f_i$, $i=0,b$, which exceed the number of
associated hydrodynamic fields:
scalar density $\rho$, (1), flow velocity $u_a$ (a=1,3) 
and momentum-flux tensor $P_{ab}$ (6), totalling 10 independent
fields. Hence, LB requires about twice as much memory as compared
to a corresponding Navier-Stokes based computational fluid dynamics method.  
This redundance buys several major computational advantages, primarily the
fact that streaming is exact and diffusion is emergent (no need of second
order spatial derivatives), which proved invaluable assets in achieving
outstanding parallel efficiency across virtually any HPC platforms, also in the
presence of real-world complex geometries \cite{}.
 
These extra-memory requirements put a premium on strategies
aimed at minimizing the cost of data access in massively parallel LB
codes \cite{}. Among others, recognized techniques are the
hierarchical access and the use of special crystallographic lattices \cite{namburi}.

In this paper we present an alternative idea, namely run the LB scheme with no
need of storing the populations.
Let's unwrap the idea in short.
Let $f=[f_1, \dots f_b]$ the $b$-dimensional array (multiscalar) 
associated with the discrete populations and $H=[\rho,u_a,P_{ab}]$ the
array associated with the hydro-fields (including the dissipative component).
The hydrodynamic fields are linear and local combinations of the
populations \cite{kruger2009shear}, that is
\begin{eqnarray}
\label{HYD}
\rho=\sum_i f_i,\\
\rho u_a=\sum_i f_i c_{ia},\\
P_{ab} = \sum_i f_i c_{ia} c_{ib}, 
\end{eqnarray}
which we write in compact form simply as $H=Pf$,
where $P$ is a projector matrix defined by the above relations.
The inverse operation, namely the reconstruction of $f$ from $H$,
reads as $f=RH$, where the reconstruction operator is the 
pseudo-inverse $P$.    
Clearly, the latter is ill-posed since the information removed
by the projector $P$ cannot be retraced back exactly.
Formally, we write $f=h+g$, where $h$ is the hydro-component of
$f$, the one that can be reconstructed exactly from $H$, whereas 
$g$ is the residual lost in the projection, known as ``ghost'' component
in LB jargon \cite{succi2018}. At zeroth order, one can simply set $g=0$.

In concrete terms, the reconstruction operator amounts to 
the following expression:\\
\begin{equation}
\label{eq:oper}
\begin{cases}
h_i = w_i \left[\rho \left( 1+ c_{ia} u_{ia} + Q_{iab} u_{a} u_{b} \right) 
+Q_{iab} (P_{ab}-P_{ab}^{eq})\right]\\
g_i = 0 
\end{cases}
\end{equation}
where $w_i$ are a set of coefficients whose values depend on the lattice geometry, $u_{ia}=u_{a}/c_s^2$, and $Q_{iab}=(c_{ia} c_{ib} - c_s^2\delta_{ab} )/(2c_s^4)$. Note that, unlike the implementation discussed in Refs\cite{chen2018highly,chen2017,chen2017simplified}, also called  simplified lattice Boltzmann method, the populations $h_i$ are reconstructed on the fly as the sum of the equilibrium and non-equilibrium terms. In particular, the first is assessed as:
\begin{equation}
\label{eq:eq-part}
h_i^{eq} =w_i \rho \left( 1+ c_{ia} u_{ia} + Q_{iab} u_{a} u_{b} \right),
\end{equation}
whereas the latter is given by 
\begin{equation}
\label{eq:neq-part}
h_i^{neq} = w_i Q_{iab}  (P_{ab}-P_{ab}^{eq}),
\end{equation}
as shown in Ref. \cite{latt2006,zhang2006efficient}.
All the higher-order non-equilibrium information, the ghost components, is discarded, i.e. $g_i=0$.

The reconstructed populations $h_i$ are then streamed and relaxed
without storing them (see next section). Note that streaming
revives the ghost components $g_i$, which affect only the non-equilibrium part of the set of populations.
As a result, the operational sequence at each time-step is
Reconstruct-Stream-Collide (RSC).
It should be emphasized that the RSC sequence does not match exactly
the result of the original, populations preserving, LB scheme.
The underlying assumption, though, is that the difference introduced
by the R-step naturally heals itself via the ghost revival described
above. Since the ghosts contain high-order, pre-hydrodynamic non-equilibrium
information, this assumption appears well justified, except perhaps in the 
case of very low viscous flows (turbulence), which lie outside the scope
of this paper. 

\section{Lattice Boltzmann without populations: implementation details}

\subsection{Implementation procedure}

The basic idea is to construct the populations on the fly, based on Eq.\ref{eq:oper} without storing them.
In order to caution against over-writing the values at time $t$ and $t+1$, two copies
of the hydrodynamic arrays must be stored, doubling the memory request. 
For single component fluids in the paradigm of Flip-Flop memory access pattern \cite{myre2011performance}, the memory saving is significant, 20 hydro-arrays (1 for density, 3 for velocity vector, 6 for the pressure tensor, two times in the Flip-Flop paradigm) versus 38 populations for the D3Q19 lattice scheme, adopted in the present investigation.
For two component fluids, the advantage increases. Indeed, adopting the color-gradient model \cite{wen2019improved,leclaire2017generalized,liu2012three}, we need to add two extra arrays for the second density component obtaining 22 hydro-arrays versus 76 populations, 38 populations per each fluid component.

For a single component, collisions are local and proceed according to the standard BGK update
\begin{equation}
\label{eq:coll}
h'_i = (1-\omega) h_i +\omega h_i^{eq},
\end{equation}
where all $h_i$ are stored as temporary scalars, thus without  arrays.

Inserting Eqs \ref{eq:oper} and \ref{eq:eq-part} into Eq. \ref{eq:coll}, we obtain:
\begin{align}
\label{eq:coll2}
h'_i &=  w_i \rho \left( 1+ c_{ia} u_{ia} + Q_{iab} u_{a} u_{b} \right)  +\nonumber\\
&+(1-\omega) w_i  Q_{iab} (P_{ab}-P_{ab}^{eq}),
\end{align}

The streaming is a little subtler because the streamed populations are
constructed on the fly from the hydrodynamics fields in the neighbor points and pulled by the neighbor sites (non-local).
In equations:
\begin{align}
\label{eq:stream}
 h_i(x,y,z;t+1) &=\sum_i h'_i[ \rho(\vec{x}-\vec{c}_i,t),u_{a}(\vec{x}-\vec{c}_i,t),P_{ab}(\vec{x}-\vec{c}_i,t)],
\end{align}
where $h'_i$ is the post-collision  population that is the function of the hydrodynamic fields $\rho$, $u_a$, and $P_{ab}$ in the neighbor lattice nodes. The collision and streaming steps can be fused into a unique step (fused approach) or treated as two steps (split approach).
In the latter case, after the collision step of Eq. \ref{eq:coll}  the post-collision hydrodynamics fields are assessed and stored in the GPU memory, while in the fused approach the updated hydrodynamics fields are assessed and stored in the GPU memory only after Eq. \ref{eq:stream}. As aforementioned, the best strategy for the fused approach could be implementing the Reconstruct-Stream-Collide (RSC) sequence, since it combines the non-locality of the pulled streaming with the locality of the collision step. In the following, we exploit the color-gradient model described in the appendix of Ref. \cite{bonaccorso2022lbcuda} to model a bi-component fluid system. Thus, we adopt the split approach that makes simpler the assessment of the local density gradients which are necessary for the color-gradient approach. Indeed, the density in all the lattice points should be assessed after the streaming step to compute the density gradients before the collision step. Then, the perturbation and re-coloring collision operators of the color-gradient model are applied to the populations obtained from Eq. \ref{eq:coll2}.     

\subsection{Near-contact interactions}

Recently, an extension of the Lattice Boltzmann method to model near contact interactions among interfaces (in the following denoted LBNCI) \cite{montessori2019mesoscale,montessori2019modeling}
has granted the possibility to simulate a broad variety of complex flow systems, such as foams, emulsions, flowing 
collections of droplets, bubbles, and high internal phase emulsions, i.e. foamy-like media displaying ordered patterns of fluid drops when under confinement \cite{bogdan2022stochastic,montessori2021translocation,montessori2021wet}. In those systems, 
the relative motion of two fluid interfaces in close contact represents a complex hydrodynamic process 
due to the concurrent action of several repulsive and/or attractive forces, such as electrostatic and Van der Waals forces, hydration repulsion, steric interactions, and depletion attraction \cite{stubenrauch2003disjoining,bergeron1999forces}.
Importantly, these forces govern the physics of phenomena, such as 
coalescence and/or repulsion between neighboring droplets that can considerably alter the mechanics of the material, occurring at lengthscales much larger than the near-contact ones
\cite{davis1989lubrication,barnocky1989lubrication,rubin2017elastic,shi1994cascade,mani2010events}.
Their nature has been initially elucidated by  
the pioneering works of Gibbs and Marangoni on the thermodynamics 
of liquid thin films \cite{bergeron1999forces} and  by the seminal contributions of Derjaguin and Overbeek \cite{derjaguin1940repulsive,verwey1947theory} 
concurring in the development of the DLVO theory. 

The LBNCI method is a variant of an LBM approach 
for multicomponent flows, based on the color-gradient model\cite{leclaire2017generalized}, augmented with an extra forcing term capturing the
effects of the near-contact forces acting at the fluid interface level. This method still holds to a continuum description of the interface dynamics,
even though near interaction forces arise at a molecular level down to nanometers (and below), namely the relevant spatial scale of contact forces. In other words, the mesoscale representation of near-contact forces provides a coarse-grained 
description able to retain computational viability without compromising 
the underlying physics.

In particular, the additional term is included within the LB method via an extra forcing contribution that acts only on the fluid interfaces in near contact, and it is given by

\begin{equation}
\vec{F}_{rep}= - A_{h}[h(\vec{x})]\vec{n} \delta_I .
\end{equation}
 
In the above Equation, $A_h[ h(\vec{x})]$ is the parameter controlling the strength (force per unit volume)
of the near contact interactions, $h(\vec{x})$ is the distance between two close interfaces,
$\vec{n}$ is a unit vector normal to the interface, 
and $\delta_I\propto \nabla\phi$ is a function, 
proportional to the phase field $\phi=\frac{\rho^1-\rho^2}{\rho^1+\rho^2}$, 
that localizes the force at the interface.

After writing the near contact force as the gradient of a potential,
$\vec{F}_{rep}= \nabla \pi$, and performing a Chapman-Enskog expansion, 
it can be shown that the actual numerical algorithm in the hydrodynamic limit 
approaches the equation of the linear momentum
\begin{equation}
\frac{\partial \rho \vec{u}}{\partial t} + \nabla \cdot {\rho \vec{u}\vec{u}}=-\nabla p + \nabla \cdot [\rho \nu (\nabla \vec{u} + \nabla \vec{u}^T)] + \vec{F}_{surf}
+\vec{F}_{rep}
\end{equation}
where $\vec{F}_{surf}=-\sigma(\nabla\cdot\vec{n})\vec{n} \delta_I $ is a volume interfacial force with $\sigma$ denoting the surface tension \cite{montessori2019amesoscale}, $p$ is the pressure and $\nu=c_s^2(\tau-1/2)$ is the kinematic viscosity of the mixture, being $\tau$ the single relaxation 
time and $c_s=1/\sqrt{3}$ the sound speed of the model \cite{succi2018,kruger2017lattice}.
This is the Navier-Stokes equation for a multicomponent system augmented 
with a surface-localized repulsive term.

\section{Validation}
\label{sec:val}
\begin{figure*}[htbp]
\includegraphics[width=1.0\linewidth]{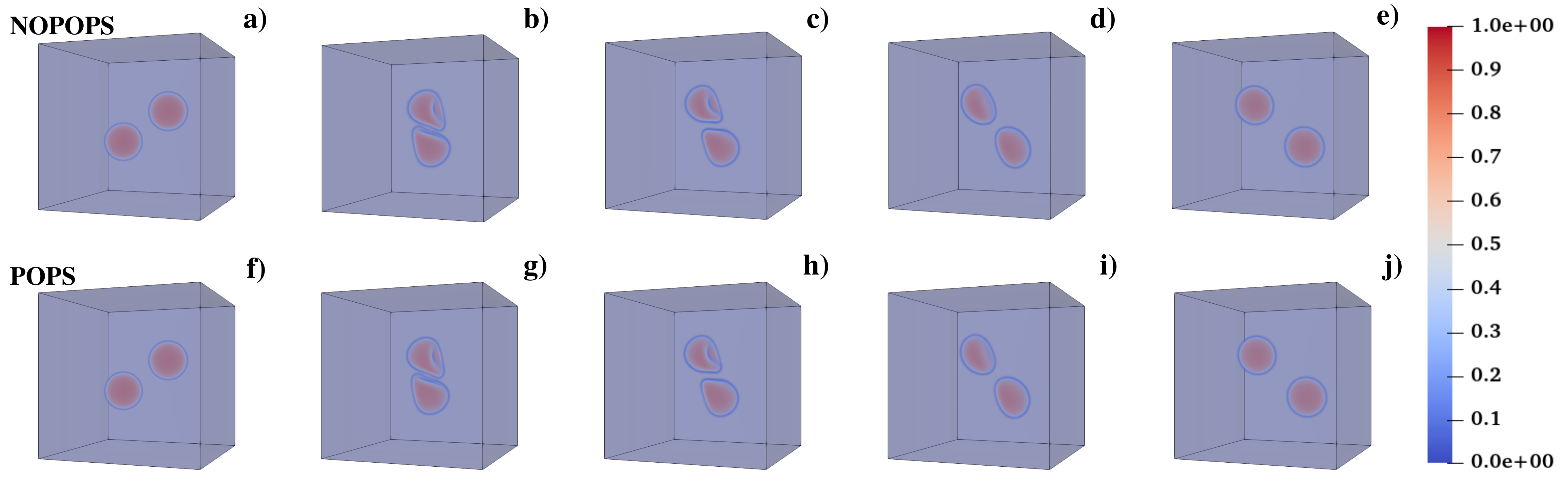}
\caption{Time sequence of an off-axis collision between two fluid droplets simulated using an LB scheme without populations (a-e) and with populations (f-j). 
The centers of mass are initially placed at a distance where the mutual interaction of the droplet is negligible. 
The two droplets, initially perfectly spherical (a)-(f), deform acquiring a bullet-like shape as they collide (b-c) and (g-h). Afterward, they separate (d)-(i) and attain a spherical shape once the mutual distance is sufficiently large (e)-(j). No appreciable differences emerge between the two approaches. The color map represents the values of the density field ranging from $0$ to $1$.}
\label{fig1}
\end{figure*}

\begin{figure}[htbp]
\includegraphics[width=1.0\linewidth]{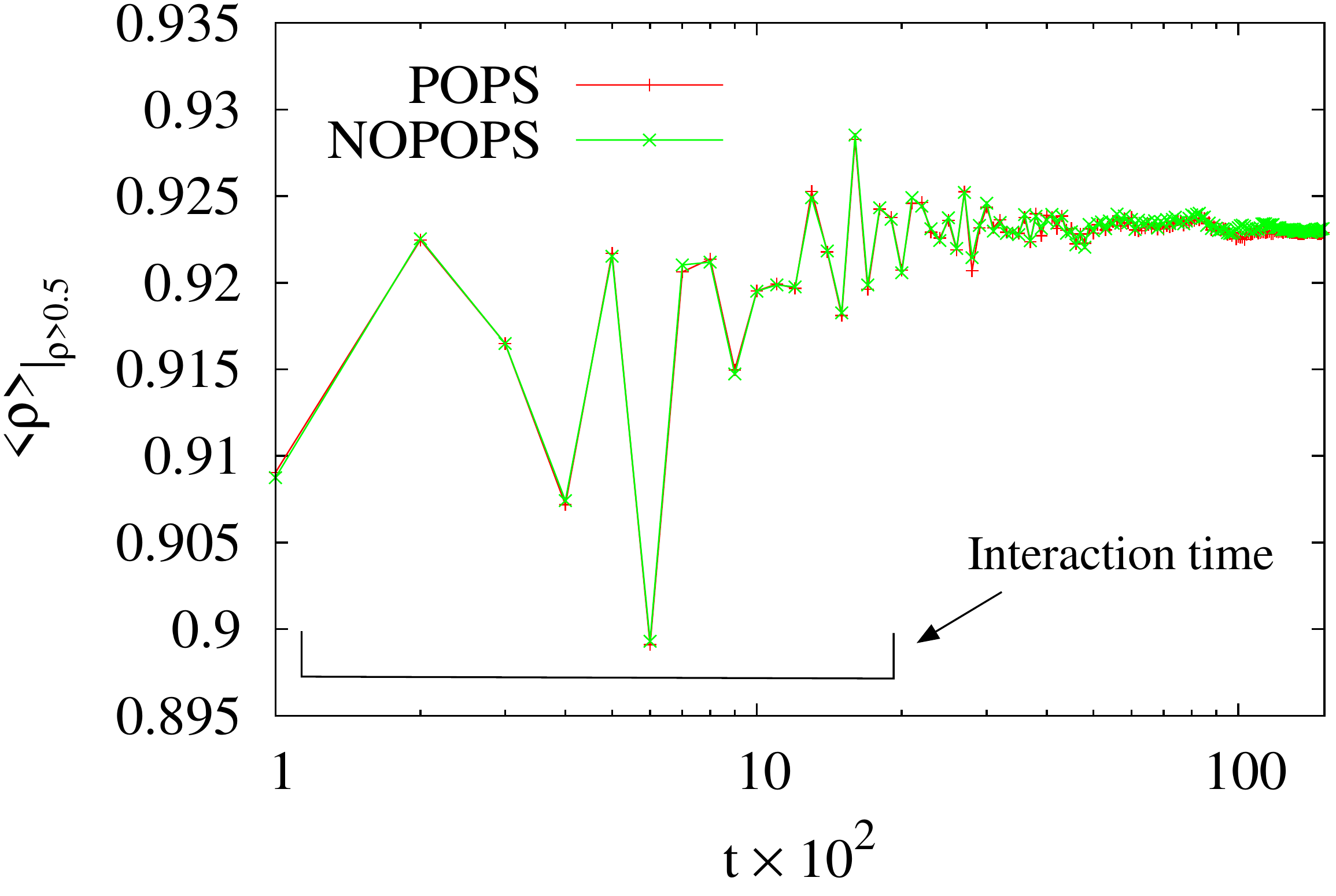}
\caption{Time evolution of $\langle\rho\rangle_{|{\rho>0.5}}$, where $\langle...\rangle$ is a spatial average computed considering lattice sites with $\rho>0.5$ (i.e. within the drops). The interaction time between the drops is $10^2 \lesssim t\lesssim 15\times 10^2$. Differences between the plots are negligible along all the evolution.}
\label{fig2}
\end{figure}

\begin{figure}[htbp]
\includegraphics[width=1.0\linewidth]{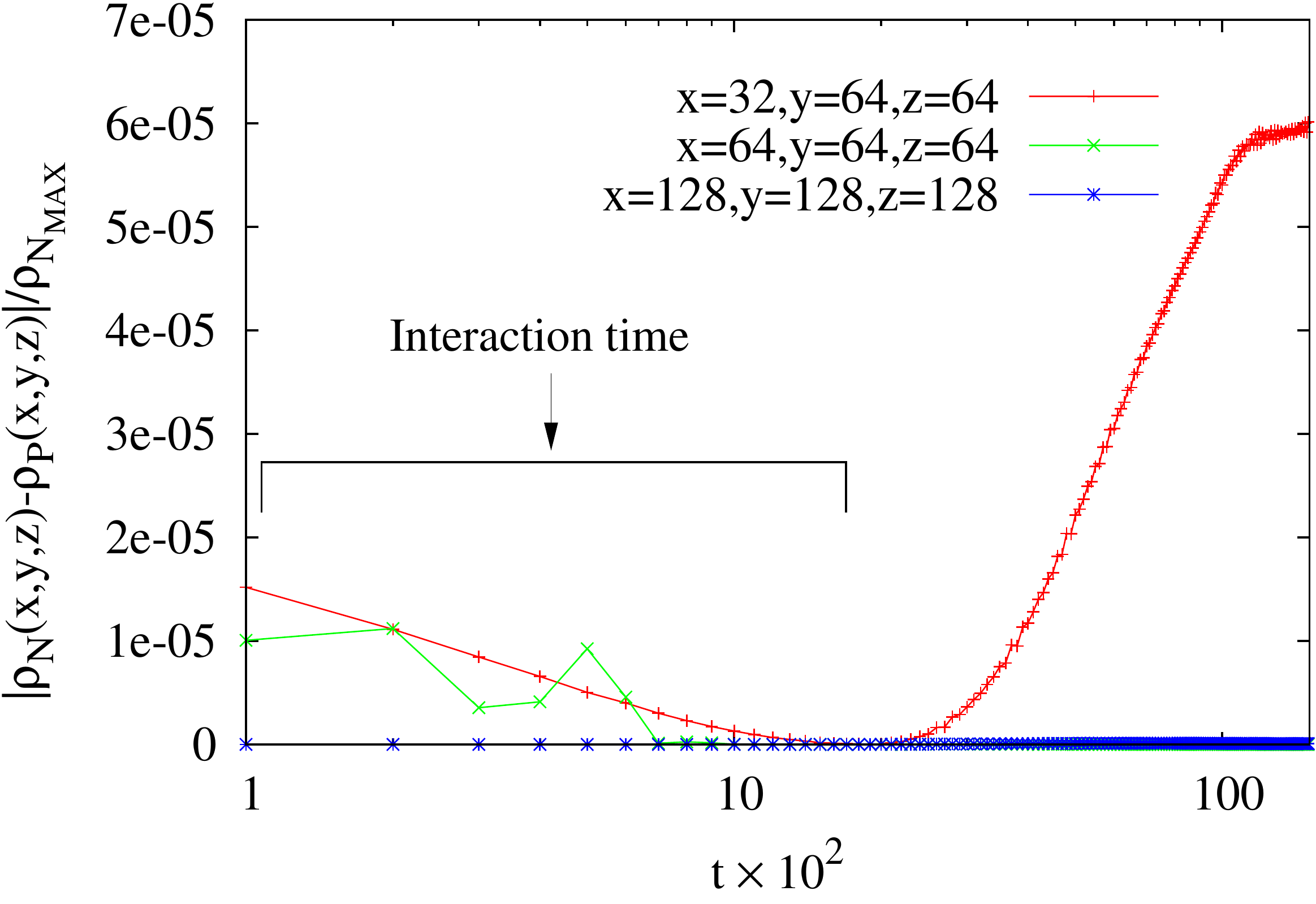}
\caption{Typical time evolution of the relative error $\frac{|\rho_{N}-\rho_{P}|}{\rho_{N_{MAX}}}$ calculated at three different sites, i.e. near the droplet interface ($x=32$, $y=64$, $z=64$), in between the drops ($x=64$, $y=64$, $z=64$) and far from the drops ($x=128$, $y=128$, $z=128$).}
\label{fig3}
\end{figure}

As a validation test, we present a simulation of a head-off-axis collision between two equally sized fluid droplets. Such a process has been extensively investigated in a previous work using a CPU-based color-gradient lattice Boltzmann method augmented with repulsive near-contact interactions preventing coalescence \cite{montessori2019mesoscale}. Here we study the collision using a GPU-based version of that method implemented with and without populations, which has been included in LBcuda \cite{bonaccorso2022lbcuda}, a GPU-accelerated version of LBsoft \cite{bonaccorso2020lbsoft}. 

\begin{figure*}[htbp]
\includegraphics[width=1.0\linewidth]{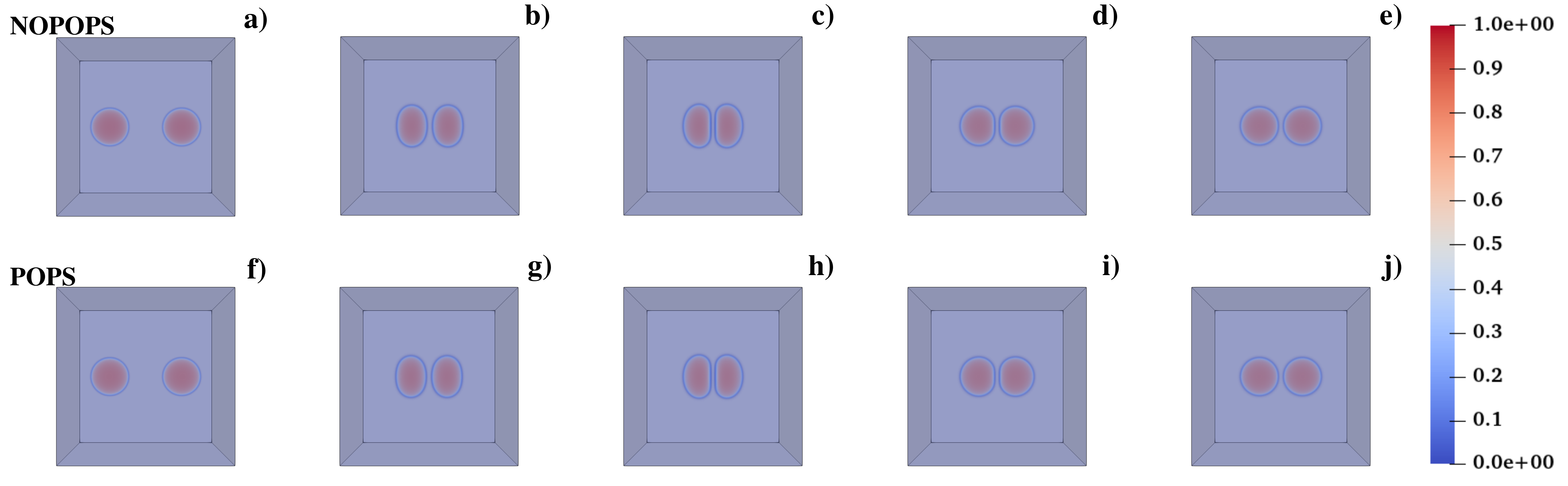}
\caption{Time sequence of a head-on collision between two fluid droplets simulated using an LB scheme without populations (a-e) and with populations (f-j). 
The centers of mass are initially placed at a distance where the mutual interaction of the droplet is negligible. 
The two droplets, initially spherical (a)-(f), 
approach and, once in close contact, elongate vertically flattening where opposite interfaces are closer, (b-c) and (g-h). Afterward, they progressively relax towards equilibrium and reacquire a spherical shape, (d-e) and (i-j).  Once again, no appreciable differences emerge between the two schemes. The color map represents the values of the density field ranging from $0$ to $1$.}
\label{fig4}
\end{figure*}

The numerical experiment is as follows. Two drops, having diameter $D=30$ lattice sites, of a fluid are dispersed in a second fluid, with a viscosity ratio fixed to one. The system is placed within a cubic box of linear size $L=128$ and the drops are located at a distance such that reciprocal interactions are negligible. The impact parameter (i.e., the distance between the collision trajectories perpendicular to their relative direction of motion) is $b=0.85$, the strength of near-contact forces preventing coalescence is $A_h=0.01$, while other values can be found in Table 1 of Ref.\cite{montessori2019mesoscale}.
In Fig.\ref{fig1} we show the time sequence of such a collision simulated using a LB approach without populations (a-e) and with populations (f-j). In both cases the relative impact velocity $U_{rel}$  is set to $0.5$. The two drops initially approach and then impact without merging (Fig.\ref{fig1}b,c and g,h), since the repulsive near-contact interactions prevent the coalescence and enable the formation of a thin film that separates the drops \cite{montessori2019mesoscale}.
During the bouncing, the shape of the drops considerably departs from the initial spherical geometry, temporarily flattening where opposite interfaces come into close contact. Afterwards, they separate and gradually reacquire a spherical shape when their reciprocal distance is sufficiently high.

The time sequence of Fig.\ref{fig1} shows that the two numerical approaches with and without populations essentially lead to equivalent dynamic behaviors. On a quantitative basis, this result is corroborated by evaluating, for example, the time evolution of the density field $\langle\rho\rangle_{|_{\rho>0.5}}$ shown in Fig.\ref{fig2}, where $\langle ...\rangle$ represents a spatial average computed over lattice points for which $\rho>0.5$ (i.e. within both drops, the red regions in Fig.\ref{fig1}). Larger fluctuations occur only at early times, i.e. when the droplets come into contact,  while they progressively shrink as the droplets move away. Note that the differences between the two plots remain marginal over the full process. In Fig.\ref{fig3} we compute, at two different lattice sites, the time evolution of the relative error
$\frac{|\rho_{N}-\rho_{P}|}{\rho_{N_{MAX}}}$, where 
$\rho_N$ and $\rho_P$ are the densities of the LB without and with populations, while $\rho_{N_{MAX}}$ is the maximum value of $\rho$ calculated using LB without populations. Unlike the previous one, this quantity provides a {\it local} measure of potential discrepancies between the two approaches. The plot shows that the relative error remains rather small over time, a further indication that the procedure of reconstructing the populations proposed in this work preserves, at a high level of accuracy, the physics of the systems. 

\begin{figure}[htbp]
\includegraphics[width=1.0\linewidth]{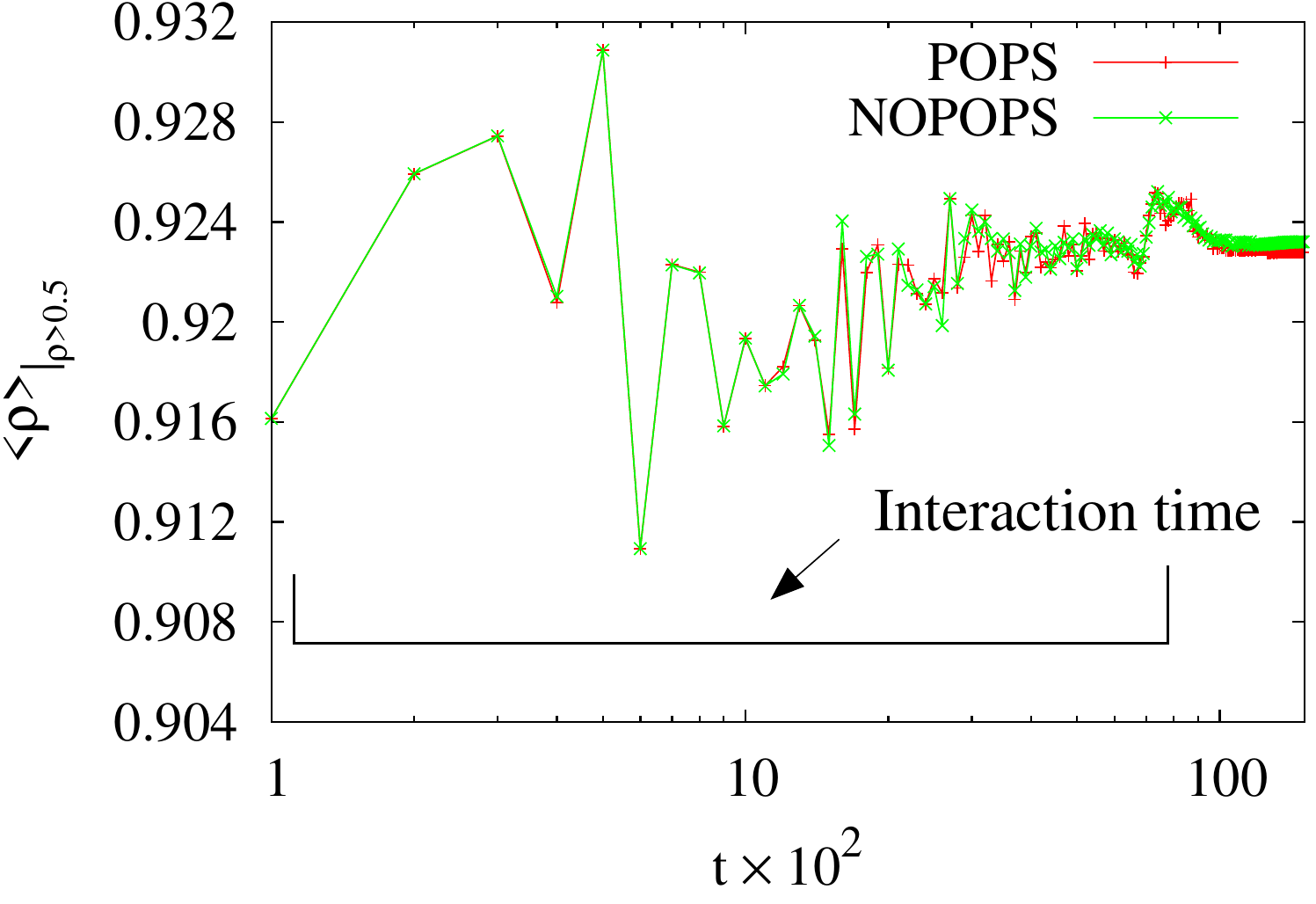}
\caption{Time evolution of $\langle\rho\rangle_{|{\rho>0.5}}$, where $\langle...\rangle$ is a spatial average computed considering lattice sites with $\rho>0.5$ (i.e. within the drops). The interaction time between the drops is $10^2 \lesssim t\lesssim 80\times 10^2$.}
\label{fig5}
\end{figure}

The approach without populations has been also tested in simulations of a head-on collision between two fluid droplets (see Fig.\ref{fig4}). Here $b=0$, $U_{rel}=0.25$ and $D=30$. Once again, the physics is correctly reproduced and no appreciable discrepancies emerge, as also shown in Fig.\ref{fig5} where the time evolution of $\langle\rho\rangle_{\rho>0.5}$ is plotted. Note incidentally that, in this case, the interaction time is considerably longer than before since, after the bouncing, the drops remain in close contact for a longer period of time.

\begin{figure*}[htbp]
\includegraphics[width=1.0\linewidth]{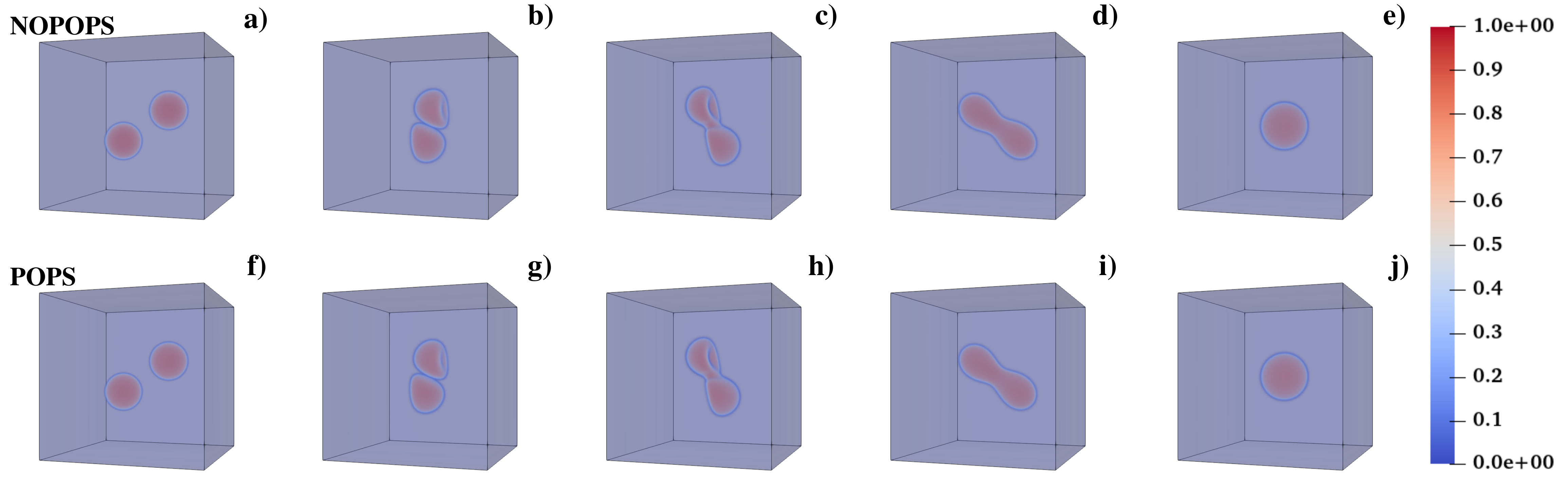}
\caption{Time sequence of the coalescence between two fluid droplets induced by a collision. Snapshots (a)-(e) show the results obtained using an LB scheme without populations while (f)-(j) the ones with populations.  The centers of mass are initially placed at a distance where the mutual interaction of the droplet is negligible.
Here the drops, initially separated (a,f), come into close contact (b,g) and connect through a thin neck of red fluid (c,h). Afterward, they slowly merge morphing into a single droplet.}
\label{fig6}
\end{figure*}

\begin{figure}[htbp]
\includegraphics[width=1.0\linewidth]{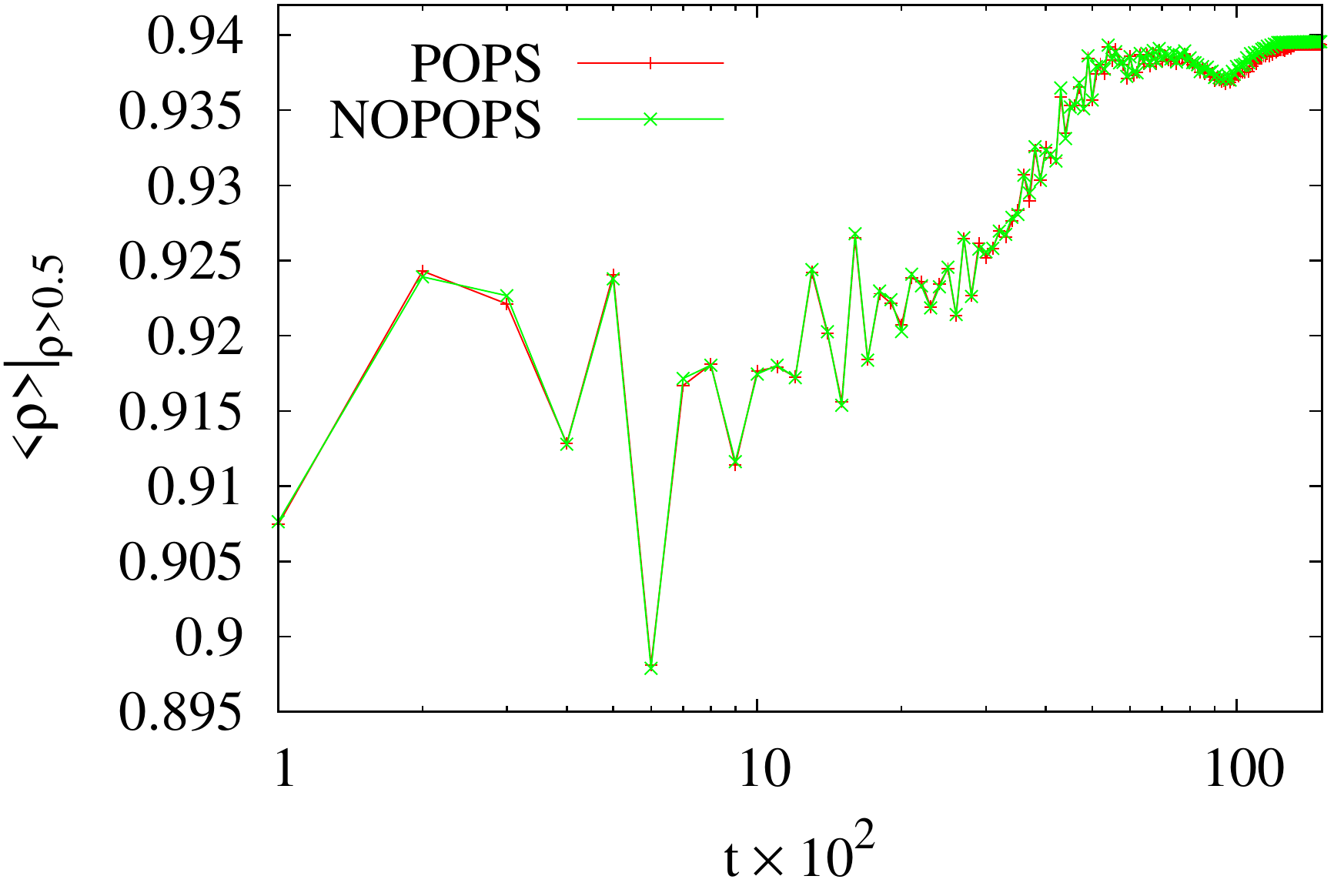}
\caption{Time evolution of $\langle\rho\rangle_{|{\rho>0.5}}$, where $\langle...\rangle$ is a spatial average computed considering lattice sites with $\rho>0.5$. Fluctuations observed for $t\lesssim 60\times 10^2$ indicate that the droplets interact and the merging is ongoing. Once again, differences between the two methods remain negligible.}
\label{fig7}
\end{figure}

The numerical experiments discussed so far present results in which near-contact interactions function as an effective force field preventing droplet coalescence. But what if these forces are turned off? In Fig.\ref{fig6} we simulate precisely this case by setting $A_h$ equal to zero (other parameters are the same as those of Fig.\ref{fig1}). The drops, initially static and perfectly spherical, approach and connect through a tiny neck of red fluid, a region in which capillary forces are strong enough to prevent its breakup. Afterward, the two droplets merge into a single one which slowly attains an equilibrated spherical shape. Once again, both methods with and without populations describe the physics of the process with negligible differences, as also demonstrated in Fig.\ref{fig7}, where early times density fluctuations capture the dynamics of droplet coalescence.

\subsection{Performance data}

To compare the performance, we use the head-off-axis collision
reported in Section \ref{sec:val} as a benchmark case.
It consists of a cubic box of 128 lattice points per side with a two-component fluid. 
All the tests were carried out on an NVIDIA® Tesla® V100 Tensor Core equipped with 16 GB of GPU memory.
We exploit the Mega Lattice Updates Per Second (MLUPS) metrics to probe the efficiency. In particular, 
the definition of MLUPS reads:
\begin{equation}
\label{eq:mlups}
\text{MLUPS}=\frac{L_x L_y L_z}{10^6 t_{\text{s}}},
\end{equation}
where $t_{\text{s}}$ is the run (wall-clock) time (in seconds) per single time step iteration, while $L_x$, $L_y$, and $L_z$ are the domain sizes.
The current version without populations runs at about 622 MLUPS against the 277 MLUPS of the standard CGLB.
A number of specific CUDA optimization techniques have been developed to improve the performance, to be detailed in a future and more technical presentation.
Briefly, the main difficulty is in improving the streaming step, which requires reading the hydrodynamic fields (11 arrays) in the 18 neighbor points for a total of 198 data accesses (792 bytes in single precision). Although the readings are scattered, neighbor threads share several readings of the hydrodynamic fields. 
As a consequence, threads in the same thread block may cooperate in the data access by using shared memory. Nonetheless, the shared memory is set to 48 Kbytes by default, and it is configurable up to 96 Kbytes per streaming multiprocess in the NVIDIA® V100 card. 
In the present work, we exploit the shared memory to store only the density fields, obtaining a performance improvement of about 20 percent. In this context, the larger size of shared memory (up to 164 Kbytes) in the NVIDIA® Ampere® A100 Tensor Core GPU will probably grant an important performance gain, allowing to store also the velocity vector in the shared memory.
Finally, the main advantage is observed in the GPU memory usage that is equal to 0.6782 GB in the version without populations compared to the version with populations where it is equal to 1.1411 GB, $\sim 40\%$ of GPU memory saving. Note that the advantage is strictly connected to the three-dimensional lattice model implemented in the code.
A 40\% memory saving is highly appealing on very large-scale simulations with multi-billion grid points since it amounts to hundreds of Gbytes savings \cite{giacomo_nat}.
More importantly, such savings are destined to become much more significant for the case of multiple species, such as those occurring in many natural and industrial applications.

\section{Summary and outlook}
In summary, we have introduced a GPU-accelerated lattice Boltzmann method that reconstructs the distribution functions from hydrodynamics quantities (density, momentum and pressure tensor) and does not require the  storage of the full set of populations. This entails a significant decrease of memory requirements and a substantial enhancement of the computational performance, features of particular relevance for large-scale simulations. The accuracy of the method has been validated in numerical experiments of relevance to soft matter, such as the collision of fluid droplets. For such cases, the formulation without populations shows satisfactory numerical stability and negligible differences with respect to a standard reconstruction-free LB approach. 
In practical terms, 40$\%$ memory reduction essentially means doubling the volume of the simulation which, for the case of a simulation demanding about  10 TB of data, would save roughly 4
TB memory \cite{giacomo_nat,muphy,keyes}.
It would be of interest to further test the model on more complex states of matter, such as colloidal fluids and fluid droplets flowing within microfluidic channels, where the implementation of solid boundaries is often necessary \cite{benzi2006}. 
On a more general basis, our results could be helpful for the design of novel high-performance computing methods in soft matter, where customized numerical schemes are fundamental to tackle phenomena occurring over a broad spectrum of scales in space and time. 

\section*{acknowledgments}
The research leading to these results has received funding from MIUR under the project ``3D-Phys'' (No. PRIN 2017PHRM8X) and from the European Research Council under the European Union's Horizon 2020 Framework Programme (No. FP/2014-2020)/ERC Grant Agreement No. 739964 (``COPMAT''). 
We acknowledge CINECA Project No. IsB25\_DIW3F under the ISCRA initiative, for the availability of high-performance computing resources and support. 
F.B. acknowledges funding by the European Research Council (ERC) under the European Union's Horizon 2020 Research and Innovation Program (Grant Agreement No. 882340). 
M.L. acknowledges the support of the Italian National Group for Mathematical Physics (GNFM-INdAM).

\section*{Data Availability Statement}
The data supporting this study's findings are available from the corresponding author upon reasonable request.


\providecommand{\noopsort}[1]{}\providecommand{\singleletter}[1]{#1}%

\end{document}